\documentclass[twocolumn,showpacs,preprintnumbers,amsmath,amssymb]{revtex4}
\usepackage{tabularx,graphicx}

\usepackage{color}
\usepackage{hyperref}
\hypersetup{
    colorlinks=true,
    linkcolor=blue,
    filecolor=blue,      
    urlcolor=blue,
}

\begin{document}

\newcommand{\beq}{\begin{equation}}
\newcommand{\eeq}{\end{equation}}
\newcommand{\beqn}{\begin{eqnarray}}
\newcommand{\eeqn}{\end{eqnarray}}
\newcommand{\bmath}{\begin{subequations}}
\newcommand{\emath}{\end{subequations}}
\newcommand{\bra}[1]{\langle #1|}
\newcommand{\ket}[1]{|#1\rangle}

\title{What is the speed of the supercurrent in superconductors?}
\author{J. E. Hirsch }
\address{Department of Physics, University of California, San Diego,
La Jolla, CA 92093-0319}

\begin{abstract} 
Within the conventional theory  of superconductivity  superfluid carriers respond to an applied magnetic field and acquire a speed according
to their effective (band) mass.  On the other hand it can be shown theoretically and is confirmed experimentally that the mechanical momentum of the supercurrent carriers is given by the product of the supercurrent speed and 
  the $bare$ electron mass. By combining these two well-established facts we show   that   the conventional
  BCS-London theory of superconductivity applied to Bloch electrons is
  {\it internally inconsistent}. Furthermore, we argue that BCS-London theory with Bloch electrons does not describe
   the phase rigidity and macroscopic quantum behavior exhibited by superconductors.
  Experimentally the speed of the supercurrent in superconductors has never been    measured and
  has been argued to be non-measurable,
  however we point out that it is in principle measurable by a Compton scattering experiment.
  We predict that such experiments will show that  superfluid carriers respond to an applied magnetic field
  according to their bare mass, in other words, that they respond  as $free$ $electrons$,
  undressed from the electron-ion interaction, rather than as Bloch electrons. This is inconsistent with the conventional theory of superconductivity
   and consistent with the alternative theory of hole superconductivity. Furthermore we point out that in principle Compton scattering
  experiments can also detect the presence of a spin current in the ground state of superconductors predicted by
  the  theory of hole superconductivity.
 \end{abstract}
\maketitle

\section{introduction}
Consider a superconducting cylinder of radius $R$ and height $h$ in an applied magnetic field $H$
smaller than the lower critical field $H_{c1}$  in direction parallel to its axis ($\hat{z}$ direction),
as shown in Fig. 1. We assume the current response to the magnetic vector potential is local and that the superconductor is in the clean limit.
An azimuthal  current $I$ flows within a London penetration depth $\lambda_L$ of the surface to nullify the magnetic field in the interior, given by
\beq
I=\frac{c}{4\pi}hH
\eeq
and the current density is given by
\beq
J=\frac{c}{4\pi\lambda_L}H
\eeq
as follows from Ampere's law $\vec{\nabla}\times\vec{H}=(4\pi/c)\vec{J}$ and the requirement that $\vec{B}=0$ inside the superconductor. 
The current density is given by
\beq
\vec{J}=n_s e \vec{v}_s
\eeq
where $n_s$ is the density of superconducting carriers and $v_s$ the superfluid velocity. From Eqs. (2) and (3) the speed of the superfluid
carriers is
\beq
v_s=\frac{c}{4\pi\lambda_L n_s e}H
\eeq

           \begin{figure}
 \resizebox{4.5cm}{!}{\includegraphics[width=6cm]{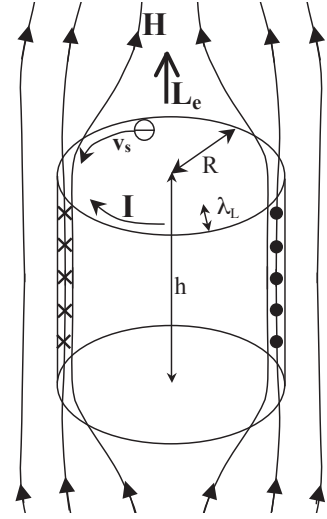}}
 \caption { Superconducting cylinder of radius $R$ and height $h$  in a magnetic field. The angular momentum of the supercurrent $\vec{L}_e$ is parallel to the
 applied field $\vec{H}$. Current $I$ flows within a shell of thickness $\lambda_L$ near the surface, in clockwise direction as seen from the top.
  }
 \label{figure1}
 \end{figure}

The London penetration depth $\lambda_L$ can be measured by standard techniques \cite{lambda,lambda2,lambda3}. In this paper we ask the question: for given
measured  values of $\lambda_L$ and $H$, what is the value of the superfluid velocity $v_s$?
It cannot be inferred from Eq. (4) because $n_s$ is not known. $\lambda_L$ determines the ratio of superfluid 
density $n_s$  to effective mass $m^*$ through the
standard relation \cite{tinkham,londonbook,schriefferbook}
\beq
\frac{1}{\lambda_L^2}=\frac{4\pi n_se^2}{m^*c^2} .
\eeq
We assume an isotropic superconductor for simplicity.
Optical \cite{optical1,optical2,optical3} as well as other experiments \cite{uemura} can measure the superfluid weight $n_s/m^*$, but there are no experiments that measure $n_s$
nor $m^*$  separately. As a consequence, it is often stated that $m^*$ for the superfluid carriers is arbitrary and at our
disposal \cite{tinkham,degennes}, for example in de Gennes' book it is stated  \cite{degennes} `We could just as well have chosen the mass of the sun', and in Tinkham's book it is stated \cite{tinkham} 
`In view of the experimental inaccessibility of $m^*$ ...'. 

In this paper we argue that this is not so. We point out that even though $v_s$ has never
been experimentally measured, it can in principle be measured through Compton scattering 
experiments \cite{compton1,compton2}. 
Theoretically, we argue   that the mass that enters in Eq. (5) is  necessarily the $bare$ electron mass $m_e$
rather than the effective (band) mass $m^*$. We show that from a purely  theoretical point of view Eq. (5) with $m^*\neq m_e$  is untenable. Since conventional BCS-London theory
predicts that $m^*$ in  Eq. (5) is the effective mass \cite{einzel,einzel2,londonours,scal,misawa,carbotte,chak,franz,gross}, which can be very different from the bare electron mass \cite{heavyf},
 this implies that    conventional BCS-London theory of superconductivity is untenable.
 On the other hand, we point out that the alternative theory of hole superconductivity \cite{holesc} predicts that 
$m^*=m_e$ in Eq. (5).

By measuring $v_s$ through Compton scattering experiments \cite{compton1,compton2}
one would obtain $n_s$ from Eq. (4) and $m^*$ 
 from Eq. (5), 
 \beq
 m^*=\frac{e\lambda_L H}{cv_s} ,
 \eeq
 hence our prediction that $m^*$ is necessarily $m_e$ can be tested experimentally.  

\section{Superfluid velocity in the Conventional theory}
In this section and in Appendix A we review   the conventional arguments from which it follows that the mass in Eq. (5) is the band effective 
mass within BCS theory \cite{einzel,einzel2,londonours,scal,misawa,carbotte,chak,franz,gross}.
 In subsequent sections we will show that this conclusion is untenable in view of experimental properties of superconductors. For simplicity we consider only zero temperature. This can 
be extended to finite temperatures along the lines of Ref. \cite{einzel}.
We assume for simplicity and definiteness that any   corrections to the band effective mass arising
from Fermi liquid effects, electron-phonon interactions, etc, can be ignored.

We consider a simple one-band model to make our point clearly. The ground state wave function in BCS theory is
given by
\beq
|\Psi>=\prod_k (\bar{u}_k+\bar{v}_k c_{k\uparrow}^\dagger c_{-k\downarrow}^\dagger) |0>
\eeq
where $k$ labels Bloch states (we omit vector labels on the $k$'s for simplicity).
The mechanical   momentum of a Bloch electron $\Psi_k(\vec{r})$ with wavevector $k$ is \cite{am}
\beq
\vec{p}_k=m_e\vec{v}_k=<\Psi_k|\frac{\hbar}{i}\vec{\nabla}|\Psi_k>=m_e\frac{1}{\hbar}\frac{\partial \epsilon_k}{\partial \vec{k}}
\eeq
where  $m_e$ is the $bare$ electron mass and $\epsilon_k$ is the band energy. Note that the electron's mechanical momentum is $not$ $\hbar\vec{k}$, nor is it $m^*v_k$. 
Following the semiclassical model of electron dynamics we assume that in the presence of slowly varying external fields electrons
can be  described by wavepackets  labeled by wavevector $k$, centered at $\vec{r}$ and spread out over many lattice constants but
of spatial extent much smaller that the wavelength of any applied fields, with velocity $v_{k}(\vec{r})$.
The semiclassical equation of motion in the presence of an external force $\vec{F}_{ext}$ is
\beq
\frac{d(\hbar \vec{k})}{dt}=\vec{F}_{ext}
\eeq
so that the time evolution of the electron mechanical  momentum is, on one hand
\beq
\frac{d\vec{p}_k}{dt}=\vec{F}_{ext}+\vec{F}_{latt}
\eeq
and on the other hand
\beq
\frac{d \vec{p}_k}{dt}=\frac{m_e}{\hbar^2}\frac{\partial^2\epsilon_k}{\partial \vec{k}\partial \vec{k}} \frac{d(\hbar \vec{k})}{dt}=  \frac{m_e}{\hbar^2}\frac{\partial^2\epsilon_k}{\partial \vec{k}\partial \vec{k}} (\vec{F}_{ext})   .
\eeq
In Eq. (10), $\vec{F}_{latt}$ is the force exerted by the lattice on the electron.
We assume an isotropic band and define 
\beq
\frac{1}{m^*_k}=\frac{1}{\hbar^2}\frac{\partial^2\epsilon_k}{\partial k^2}
\eeq
Under application of a magnetic field to the cylinder the Faraday electric field that develops
within a London penetration depth of the surface is
\beq
\vec{E}_F=-\frac{\lambda_L}{c}\frac{\partial H}{\partial t}\hat{\theta}
\eeq
in the azimuthal direction, exerting an external force $\vec{F}_{ext}=e\vec{E}_F$ on electrons, so that
their mechanical momentum changes according to equation (11) as
\beq
\frac{d\vec{p}_k}{dt}=\frac{m_e}{m^*_k} (e\vec{E}_F)=-\frac{m_e}{m^*_k}  \frac{e\lambda_L}{c}\frac{\partial H}{\partial t}\hat{\theta}
\eeq
and the change in mechanical momentum and
 velocity of the electron when the magnetic field increases from $0$ to $H$ is
\bmath
\beq
\Delta \vec{p}_k= - \frac{m_e}{m^*_k}  \frac{e\lambda_L}{c}H  \hat{\theta}
\eeq
\beq
\Delta \vec{v}_k= - \frac{1}{m^*_k}  \frac{e\lambda_L}{c}H  \hat{\theta}
\eeq
\emath
respectively.
At zero temperature the occupancy of the Bloch state with wavevector $k$ is
\beq
n_k=2|\bar{v}_k|^2=2(1-|\bar{u}_k|^2) 
\eeq
where $\bar{v}_k$, $\bar{u}_k$ are the BCS amplitudes in Eq. (7), which are  $1$ or $0$  except
in the neighborhood of the Fermi surface. The current density that develops is then
\beq
\vec{J}=\frac{e}{V}\sum_{k} n_k (\vec{v}_k+\Delta \vec{v}_k)= - \frac{1}{V} \sum_{k} n_k \frac{1}{m^*_k}  \frac{e^2\lambda_L}{c}H \hat{\theta}
\eeq
since the first term gives zero. Using eq. (2) yields for the penetration depth
\beq
\frac{1}{\lambda_L^2}=\frac{4\pi e^2}{c^2}(\frac{1}{V}\sum_{k} n_k \frac{1}{m^*_k}).
\eeq
For a band that is close to empty we can assume that $m^*_k \sim m^*$  approximately independent of $k$
for the states for which $n_k\neq 0$, and Eq. (18) is
\beq
\frac{1}{\lambda_L^2}=\frac{4\pi n_s e^2}{ m^* c^2}
\eeq
since
\bmath
\beq
\frac{1}{V} \sum_k n_k \frac{1}{m^*_k}=\frac{n_s}{m^*}
\eeq
with
\beq
n_s=\frac{1}{V} \sum_{k} n_k
\eeq
\emath
the number of superfluid carriers per unit volume. The velocity shift Eq. (15b) is independent of $k$ and given by
\bmath
\beq
\Delta \vec{v}_k\equiv \vec{v}_s= - \frac{e\lambda_L}{m^*c}H   \hat{\theta}   .
\eeq
$v_s$ is the speed of the supercurrent carriers, and the supercurrent Eq. (17)  is 
\beq \vec{J}=e n_s \vec{v}_s . \eeq \emath
Similarly for a band that is close to full we use that
\beq
\sum_{all \  k}\frac{1}{m^*_k}=0
\eeq
and defining $m^*=-m^*_k$, assumed independent of $k$ for the states for which $(2-n_k)\neq 0$ we have
\bmath
\beq
\frac{1}{V}\sum_k n_k \frac{1}{m^*_k}=\frac{1}{V}\sum_k(2-n_k)\frac{1}{(-m^*_k)}=\frac{n_s}{m^*}
\eeq
with the superfluid density now given by
\beq
n_s=\frac{1}{V} \sum_{k} (2-n_k) 
\eeq
\emath
so that the same expression Eq. (19) for the London penetration depth results. 
The velocity of the supercurrent carriers is still given by Eq. (21a), and the same expression for the
supercurrent  $\vec{J}=e n_s \vec{v}_s$   results. These results can also be derived by using the standard
linear response theory formalism as discussed in Appendix A.

The fact that we end up with the band effective mass  $m^*$ in the expressions for the London penetration depth
Eq. (19) and superfluid velocity Eq. (21a) can be traced back to the form of the BCS wavefunction Eq. (7). In 
particular to the fact that within BCS theory the states $k$ are the same as in the normal metal, only
a slight change in occupation of those states occurs within a region $\Delta$ of the Fermi energy, with 
$\Delta$ the energy gap. The same results Eqs. (19) and (21a) would of course apply to a perfect conductor
rather than a superconductor.  This implies that if $m_e$ rather than $m^*$ has to appear in Eqs. (19) and (21a)
some rather  profound modification of the BCS wavefunction would be needed.

\section{mechanical momentum of the supercurrent}
The mechanical momentum of a Bloch electron is given by Eq. (8), so when the supercurrent is generated the change
in the momentum of one electron is
\beq
\Delta \vec{p}_k=m_e \vec{v}_s =-\frac{m_e}{m^*}\frac{e\lambda_L}{c}H\equiv \vec{p}_{mech}
\eeq
where we used Eq. (21a) for the superfluid velocity. 
The mechanical  momentum density per unit volume (which is zero in the absence of current) is
\bmath 
\beqn
\mathcal{P}_{mech}&=&\frac{1}{V}\sum_k n_k \Delta p_k=n_s\Delta p_k=n_s m_e v_s   \\ \nonumber
&=& \frac{m_e}{e}J=-m_e n_s \frac{e\lambda_L}{m^*c}H 
=-\frac{m_ec}{4\pi \lambda_L e}H
\eeqn
or alternatively
\beqn
\mathcal{P}_{mech}&=&\frac{1}{V}\sum_k n_k \Delta p_k= \frac{m_e}{e}J=     \\ \nonumber
&=&-\frac{m_e e \lambda_L}{c} H \frac{1}{V}\sum_k\frac{n_k}{m^*_k}=
-\frac{m_ec}{4\pi \lambda_L e}H
\eeqn
\emath
where we have used Eqs. (19) or (18) in the last equality.  Note that Eqs. (24) and (25a) apply to the particular cases where the band is close to empty or close to full,
while Eq. (25b) is valid for any band filling
The same results are obtained using the linear response
formalism in Appendix A. Note that the mechanical momentum density is independent of the
effective mass. The total angular momentum of the supercurrent for the cylinder of radius $R$ and height
$h$ is the volume of the shell of thickness $\lambda_L$ where the supercurrent flows times the momentum density
times the radius R, under the assumption that $R>>\lambda_L$:
\beq
L_e=(2 \pi R \lambda_L h)\mathcal{P}_{mech}R=-\frac{m_ec}{2e}hR^2H  .
\eeq
The total mechanical angular momentum $L_e$  is independent of $\lambda_L$, $m^*$ and $n_s$. Hence from measurement of
$L_e$ we cannot determine whether it is $m^*$ or $m_e$ that enters the equations, since $m^*$ does not enter
in Eq. (26). $L_e$ is measured experimentally in the gyromagnetic effect \cite{gyro1,gyro2,gyro3,gyro4} and its value
is found to be precisely as given by Eq. (26), which unfortunately says nothing new. It confirms however that the mechanical momentum
of the electrons carrying  the supercurrent is given by $m_ev_s$ and $not$ by $m^*v_s$.

\section{Canonical momentum of the supercurrent}
For the superconducting cylinder under consideration the relation between magnetic field and magnetic
vector potential is simply
\beq
A=\lambda_L H
\eeq
as follows  from the relation $\vec{\nabla}\times\vec{A}=\vec{H}$.
$\vec{A}$ points in the azimuthal
direction.
Eq. (27) assumes the Coulomb gauge $\vec{\nabla}\cdot\vec{A}=0$, or equivalently that  $A$ is constant along the
circumference of the cylinder.  In terms of $\vec{A}$, the mechanical momentum of a carrier of the supercurrent is, from Eq.  (24)
\beq
\vec{p}_{mech}=m_e \vec{v}_s =-\frac{m_e}{m^*}\frac{e}{c}\vec{A} .
\eeq
Now the canonical momentum $\vec{p}$    that enters the Schr\"odinger equation for a particle of 
mass $m$ and charge $q$ moving with velocity $\vec{v}_s$ in the presence of a vector potential $\vec{A}$ is
\beq
\vec{p}=m \vec{v}_s +\frac{q}{c}\vec{A}
\eeq
where $m\vec{v}_s$ is the mechanical (or `kinematic') momentum, that equals the canonical momentum
when $\vec{A}=0$. 
For a superconductor it is assumed that Eq. (29) applies with $q=2e$ and mass  $m=2m^*$, with $m^*$ the effective 
mass \cite{tinkham,london1948,laue1948,londonbook}:
\beq
\vec{p}=2m^* \vec{v}_s +\frac{2e}{c}\vec{A}=\hbar \vec{\nabla}\varphi  .
\eeq
In Eq. (30), $\varphi$ is the phase of the macroscopic wavefunction
 $\Psi_S(\vec{r})$ 
describing the superfluid \cite{merc}, given by
\beq
\Psi_S(\vec{r})=(n_s/2)^{1/2}e^{i\varphi(\vec{r})} .
\eeq
The right-hand side of Eq. (30) results from applying the momentum operator $\vec{p}=(\hbar/i)\vec{\nabla}$ to $\Psi_S(\vec{r})$   assuming the superfluid density $n_s$ is uniform in space.

We next discuss the consequences of the phase equation Eq. (30) for 
(i) flux quantization, (ii) Meissner effect, and (iii) mechanical momentum:

(i) Flux quantization: in a superconducting ring, integration of Eq. (30) along a closed path in the interior of a ring where
there is no current ($v_s=0$) leads to \cite{londonbook}
\beq
\oint\hbar\vec{\nabla}\varphi \cdot \vec{dl} =nh=\frac{2e}{c}\oint\vec{A} \cdot \vec{dl} =\frac{2e}{c}\phi
\eeq
with $n$ an integer and $\phi$ the magnetic flux, hence $\phi=n\phi_0$ with $\phi_0=hc/(2e)$ the flux quantum. 
This is verified experimentally \cite{fluxq1,fluxq2}.

(ii) Setting the canonical momentum $p=0$, as appropriate for the Meissner effect, yields for Eq. (30)
\beq
v_s=-\frac{e}{m^*c}A=-\frac{e\lambda_L}{m^* c}H
\eeq
using Eq. (27), in agreement with Eq. (21a) for the speed of the Meissner current.

(iii) Setting $A=0$ in Eq. (30) should give the mechanical momentum of a pair for the canonical momentum, hence
twice the mechanical momentum for one of the components of the pair:
\beq
p=2m^* v_s=2p_{mech}
\eeq
hence
\beq
p_{mech}=m^*v_s=   -\frac{e}{c}A=-\frac{e\lambda_L}{c}  H
\eeq
where we have used Eq. (21a) for the supercurrent velocity,
or equivalently set $\vec{\nabla}\varphi=0$ in Eq. (30). However, Eq. (35) is wrong since it contradicts
Eq. (24), and as a consequence it contradicts the results of the
gyromagnetic experiments \cite{gyro1,gyro2,gyro3} which were shown in Sect. II to be consistent
with Eq. (24), hence inconsistent with Eq. (35). 

To reiterate this crucial point: to the extent  that the BCS superfluid can be described by a 
macroscopic wavefunction $\Psi_S(\vec{r})$ with an amplitude and a phase as given by
Eq. (31), as evidenced by multiple experiments \cite{merc}, the
mechanical momentum density of the supercurrent when $\vec{A}=0$ has to be given by,
according to Eqs. (30), (31) and (33)
\bmath
\beq
\mathcal{P}_{mech}=<\Psi|\frac{\hbar}{i}\vec{\nabla}|\Psi>=\frac{n_s}{2}\hbar \vec{\nabla}\varphi=n_s m^* \vec{v}_s
=-\frac{n_se\lambda_L}{c}H .
\eeq
Eq. (36a) would yield for the total mechanical angular momentum instead of Eq. (26), using Eq. (19)
for the penetration depth
\beq
L_e=(2 \pi R \lambda_L h)\mathcal{P}_{mech}R=-\frac{m^*c}{2e}hR^2H  
\eeq
\emath
which disagrees with experiment \cite{gyro1,gyro2,gyro3}
that establishes that the mechanical angular momentum is given by Eq. (26), i.e. Eq. (36b) but
with $m_e$ rather than $m^*$, to an accuracy better than $1 \%$ \cite{gyro3}
(see footnote \cite{footnote}).

There is no way to `fix' Eq. (30) to make it consistent with Eq. (24). If we write instead of Eq. (30)
\beq
\vec{p}=2m_e \vec{v}_s +\frac{2e}{c}\vec{A}=\hbar \vec{\nabla}\varphi
\eeq
we will satisfy (i) (flux quantization) and (iii) (mechanical momentum) but obtain for (ii), i.e. setting A=0
\beq
v_s=-\frac{e}{m_ec}A=-\frac{e\lambda_L}{m_ec}H
\eeq
in contradiction with Eq. (21a). Finally, if we write instead of Eq. (30)
\beq
\frac{m^*}{m_e}\vec{p}=2m^* \vec{v}_s +\frac{2e}{c}\vec{A}=\frac{m^*}{m_e}\hbar \vec{\nabla}\varphi
\eeq
we will satisfy (ii) and (iii) but fail to satisfy (i), i.e. the flux quantum would depend on the ratio of bare mass to effective
mass, in contradiction with experiment. Or in other words, Eq. (39) violates gauge invariance.

These considerations show that the conventional BCS theory of superconductivity applied to
Bloch electrons leads to inconsistent results, in contradiction with what is generally
believed \cite{einzel,einzel2,londonours,scal,misawa,carbotte,chak,franz,gross}.
It is impossible to compatibilize the superfluid velocity Eq. (21a) depending on the effective (band) mass with the
requirements imposed by flux quantization and gauge invariance and the vast experimental evidence in favor
of a macroscopic superconducting wavefunction $\Psi_S(\vec{r})$ \cite{merc} that has   `phase rigidity',
so that $\vec{\nabla}{\varphi}=0$ in a simply connected  sample in the presence of $\vec{A}$,
which leads to the Meissner effect.
In Appendix B we present these arguments in a concise alternative form leading to the same conclusion.

We propose that to resolve this inconsistency
it is necessary to assume that the expression Eq. (21a) for the superfluid velocity is incorrect, and that the
correct expression is
\beq
v_s=-\frac{e\lambda_L}{m_ec}H=-\frac{e}{m_ec}A .
\eeq
which is consistent with Eq. (37) rather than Eq. (30) for the relation between canonical momentum
and superfluid velocity. It should also be pointed out that Eq. (37) is consistent with experiments by
Zimmermann and Mercereau \cite{zimm} and Parker and Simmonds \cite{parker}
that measured the Compton wavelength of electrons in 
Josephson junctions, and D. Scalapino  presents theoretical arguments for the validity of
Eq. (37) in ref. \cite{scal2}. 
An experiment by Jaklevic et al \cite{jak} detecting phase modulation 
by the superfluid velocity does not yield  information to decide between Eqs. (30) and (37)
without additional assumptions. 

This then  raises the questions: what was wrong in the straightforward  derivation leading to Eq. (21a),
or in the alternative equivalent derivation in Appendix A? How is Eq. (40) consistent with conventional BCS-London theory
and Bloch's theory of electrons in metals? We return to these questions in  later sections.

\section{The London moment}
The importance of the canonical momentum of superconducting electrons was already realized
 by F. London \cite{londonbook}, before BCS and  before the development of Ginzburg-Landau theory.
London introduced the `local mean value of the momentum vector of the superelectrons' $\vec{p}_s$:
\beq
\vec{p}_s=m_e\vec{v}_s+\frac{e}{c}\vec{A}=\vec{\nabla}\chi
\eeq
with $\chi$ the `superpotential', which we now would call $\hbar \varphi /2$. London deduced that
the right-hand-side of Eq. (41) is the gradient of  a scalar function from the Meissner effect, then
proceeded to predict flux quantization from this equation \cite{londonbook}. In addition, he argued that
for a superconductor rotating with angular velocity $\vec{\omega}$ one has
\beq
\vec{v}_s(\vec{r})=\vec{w}\times\vec{r}
\eeq
and a uniform magnetic field $\vec{B}$ gives rise to a magnetic vector potential
\beq
\vec{A}=\frac{\vec{B}\times\vec{r}}{2}  .
\eeq
Substitution of Eqs. (42) and (43) in Eq. (41) yields (for $\chi=0$ as appropriate for a 
simply-connected body \cite{londonbook})
\beq
(m_e\vec{\omega}+\frac{e}{2c}\vec{B})\times\vec{r}=0
\eeq
which predicts a uniform magnetic field in the interior of a superconductor rotating
with angular velocity $\vec{\omega}$
\beq
\vec{B}=-\frac{2m_e c}{e}\vec{\omega}
\eeq
as experimentally measured \cite{hild}. The fact that the experimentally measured magnetic
field is given by Eq. (45) with the bare electron mass $m_e$ confirms that the mass in Eq. (41)
has to be $m_e$ rather than the effective mass $m^*$ as in Eq. (30).
Thus, the observed magnetic field of rotating superconductors provides further 
experimental evidence
for the incorrectness of 
the BCS phase equation Eq. (30) that has  $m^*$ in place of $m_e$.

\section {The macroscopic superfluid wavefunction  and phase rigidity}
 A large number of experiments with superconductors, particularly involving Josephson junctions and
 weak links, can be understood and described by the assumption that there exists a macroscopic
 single-particle-like wavefunction
 \beq
 \Psi_S(\vec{r})=(\frac{n_s}{2})^{1/2}e^{i\varphi(\vec{r})}
 \eeq
 that describes the Cooper pair condensate \cite{merc}. It is generally assumed that Eq. (46) follows from BCS
 theory, where the phase $\varphi$ for a spatially uniform situation is given by
 \beq
 \frac{\bar{v}_k}{\bar{u}_k}=| \frac{\bar{v}_k} {\bar{u}_k}   |e^{i\varphi}
 \eeq
 with $\bar{v}_k, \bar{u}_k$ the amplitudes in the BCS wavefunction Eq. (7). However this has
 never been shown theoretically in a rigorous way \cite{rogovin}.
 
 Assuming the  phase equation Eq. (30)  is valid as required for the Meissner effect  implies that 
 \beq
<\Psi_S(\vec{r}) |\frac{\hbar}{i}\vec{\nabla}|\Psi_S(\vec{r)}>=n_s \hbar \vec{\nabla}\varphi =m^* n_s \vec{v}_s +n_s \frac{e}{c}\vec{A}  .
\eeq
In a simply connected superconductor the phase $\varphi$ is assumed to be uniform and not affected
by the application of a vector potential $\vec{A}$. This is termed the `phase rigidity' of the wavefunction.
Hence the left-hand side of Eq. (48), the expectation value of the canonical momentum of
the superfluid,  vanishes  and
this  implies the Meissner effect. More generally, Eq. (48) implies that superfluid flow is 
irrotational \cite{londonbook,schriefferbook,tinkham}.

Now in the many-body framework of BCS theory, the canonical momentum operator 
$(\hbar/i)\vec{\nabla}$ in Eq. (48) corresponds to what we call the `paramagnetic' momentum density operator in Appendix A,  given by
\beq
\mathcal{\vec{P}}_1=\sum_i \frac{\hbar}{i}\vec{\nabla}_i
\eeq
in first quantized form. We 
 show in Appendix A  that the expectation value of this operator (in second quantized form) 
 with the many-body BCS wavefunction in the presence of a 
vector potential $\vec{A}$ is
  \beq
 <\Psi|\vec{\mathcal{P}}_1|\Psi>=\frac{e}{c} \frac{1}{V}\sum_k n_k(1-\frac{m_e}{m^*_k})\vec{A}
 \eeq
which is $not$ zero in a simply connected superconductor subject to a magnetic field.
Comparing Eq. (50) with Eq. (48) we have to conclude that Eq. (48) is invalid. In other words,
 the generally held belief
that BCS theory   is consistent with `London rigidity' 
so that the left-hand side of Eq. (48) does not change under application of a weak slowly varying magnetic
field is invalid. The curl of the canonical momentum density Eq. (50) is non-zero, and therefore 
it cannot be said that within BCS theory the superfluid flow is irrotational, as generally
assumed \cite{tinkham,schriefferbook}.

From Appendix A Eq. (A22) we deduce that within BCS theory
 \beq
 <\Psi|\vec{\mathcal{P}}_1|\Psi>=\frac{m_e}{e}\vec{J}- <\Psi|\vec{\mathcal{P}}_2|\Psi>
 \eeq
 or equivalently using Eq. (A23)
 \beq
 \frac{e}{c} \frac{1}{V}\sum_k n_k(1-\frac{m_e}{m^*_k})\vec{A}=\frac{m_e}{e}\vec{J}+
  \frac{e}{c} \frac{1}{V}\sum_k n_k\vec{A}
 \eeq
 Eq. (52) yields the correct current $\vec{J}$ in a simply connected geometry, where both
 $\vec{J}$ and $\vec{A}$ go to zero in the interior of the material. However, applied to a ring
 of thickness larger than the London penetration depth, it also predicts that $\vec{A}=0$ where
 $\vec{J}=0$, which is incorrect. We conclude that 
 BCS theory does not predict flux quantization, and 
 London would not have been able to infer the
 existence of a `superpotential' \cite{londonbook,london1948} and predict
 flux quantization from Eq. (41) had he known about the BCS wavefunction.

 Note also  we can rewrite Eq. (30)  
 using $\vec{J}=n_s e\vec{v}_s$  as
\beq
<\Psi_S(\vec{r}) |\frac{\hbar}{i}\vec{\nabla}|\Psi_S(\vec{r)}>=\frac{m^*}{e}  \vec{J} +n_s \frac{e}{c}\vec{A} .
\eeq
The `many body' version of Eq. (53) would be within BCS theory
\beq
<\Psi  |\mathcal{\vec{P}}_1|\Psi>=\frac{m^*}{e}  \vec{J} +\frac{e}{c} \frac{1}{V}\sum_k n_k\vec{A}
\eeq
which yields using Eq. (50)
\beq
\vec{J} =-\frac{e^2}{c}\frac{m_e}{m^*}\frac{1}{V}\sum_k n_k(\frac{1}{m^*_k})\vec{A}
\eeq
which disagrees with the result predicted by BCS theory Eq. (A20).
Therefore, the phase equation Eq. (30) is inconsistent with BCS theory.

 If instead of the BCS phase equation Eq. (30)
 we assume that Eq. (37) is valid following   \cite{zimm,parker,jak,hild,scal2}, it implies that 
 \beq
<\Psi_S(\vec{r}) |\frac{\hbar}{i}\vec{\nabla}|\Psi_S(\vec{r)}>=n_s \hbar \vec{\nabla}\varphi =m_e n_s \vec{v}_s +n_s \frac{e}{c}\vec{A}  .
\eeq
and comparing Eq. (50) with Eq. (56) 
we have to conclude that Eq. (56) is invalid.  

 In summary, in this section we have shown in detail that the BCS formalism
 applied to Bloch electrons  is incompatible
 with the existence of a macroscopic single-particle-like superfluid
 wavefunction $\Psi_S(\vec{r})$ with a well-defined macroscopic phase
 $\varphi(\vec{r})$ that obeys either Eq. (30) or Eq. (37).

\section{kinetic energy of the supercurrent}
Consideration of the kinetic energy of carriers of the supercurrent furnishes another $independent$ argument for the
incorrectness of Eq. (21a) for the superfluid velocity.  

The kinetic energy density of the supercurrent is given by
\beq
\mathcal{K}=\frac{H^2}{8\pi}
\eeq
This follows from general arguments \cite{tinkham}, and furthermore it is a $necessary$ condition for the existence
of equilibrium between a normal and a superconducting phase   when $H$ is the critical field  \cite{londonh}. Hence the kinetic energy per carrier is
\beq
\epsilon_{kin}=\frac{1}{n_s}\frac{H^2}{8\pi}
\eeq
Replacing $H$ in terms of $v_s$ from Eq. (21a) yields
\beq
\epsilon_{kin}=\frac{1}{8\pi n_s}(\frac{m^*c}{e\lambda_L})^2 v_s^2
\eeq
and using Eq. (19) for $\lambda_L$ yields
\beq
\epsilon_{kin}=\frac{1}{2}m^* v_s^2  .
\eeq
Eq. (60) is generally assumed to be the correct expression for the kinetic energy of the  supercurrent carriers \cite{tinkham}.

However, the change in the kinetic energy of a Bloch electron when an external
field  is applied is given by
\beqn
\Delta \epsilon_{kin} &=&<\Psi_{k+\Delta k}|     -   \frac{\hbar^2}{2m_e}  \nabla^2 |\Psi_{k+\Delta k}> \\ \nonumber
&-&<\Psi_{k}|     -   \frac{\hbar^2}{2m_e}  \nabla^2        |\Psi_k>
\eeqn
with $\Delta k = e\lambda_L/ (\hbar c)  H$. Eq. (61)  is not equal to Eq. (60).
In particular there is absolutely $no$ physical basis for having  $m^*$ as prefactor in Eq. (60).   $m^*$ describes the response of Bloch electrons to external fields but is in no way associated with
the kinetic energy acquired  by the electron.

To calculate the extra kinetic energy of a carrier in the supercurrent, consider the work done on a superfluid electron labeled by wavevector $k$ under the influence of a force $\vec{F}$:
\beq
dW_k=\vec{F}\cdot d\vec{x}_k .
\eeq
where $d\vec{x}_k$ is the displacement of this wavepacket.
When a magnetic field is applied, the force is the sum of the external force originating in the Faraday electric  field and the
force exerted by the lattice on the electron, $\vec{F}_{latt}$
\beq
dW_k=(e\vec{E}_F+\vec{F}_{latt})\cdot d\vec{x}_k 
\eeq
From the semicassical equation of motion we have
\beq
m_e\frac{d\vec{v}_k}{dt}=e\vec{E}_F+\vec{F}_{latt}
\eeq
hence
\beq
dW_k=m_e\frac{d\vec{v}_k}{dt}\cdot d\vec{x}_k=m_e\vec{v}_k\cdot d\vec{v}_k=\frac{1}{2}m_e d(v_k^2)
\eeq
where we have used that for this electron 
\beq
\frac{d\vec{x}_k}{dt}=\vec{v}_k  .
\eeq
In addition we have assumed that no work is spent in changing the potential energy of the carrier.
Therefore we deduce from Eq. (65) 
\beq
dW_k=d(\frac{1}{2}m_e v_k^2)
\eeq
The velocity change is given by $\vec{v}_k\rightarrow \vec{v}_k+\vec{v}_s$, hence the change in the right-hand-side of Eq. (67) 
when work $W_k$ is done on the carrier is
\beq
W_k= \frac{1}{2}m_e (\vec{v}_k+\vec{v}_s)^2-\frac{1}{2}m_e v_k^2=
m_e \vec{v}_k\cdot \vec{v}_s+\frac{1}{2}m_e v_s^2
\eeq
The total work per unit volume is obtained by summing Eq. (68) over all  states multiplied by the occupation of each state
\beq
W\equiv \frac{1}{V}\sum_k n_k W_k=n_s\frac{1}{2}m_e v_s^2
\eeq
since the sum over the first term in Eq. (68) gives zero. Eq. (69) implies, by energy conservation, that the correct expression for the 
kinetic energy of a carrier in the supercurrent is
\beq
\epsilon_{kin}=\frac{1}{2}m_e v_s^2
\eeq
rather than Eq. (60). Eq. (70)  is consistent with the correct formula for the mechanical momentum Eq. (24). 

Now if we compute the kinetic energy density of the supercurrent using Eq. (70) for the
kinetic energy of a carrier, Eq. (21a) for the superfluid velocity, Eq. (19) for the London penetration depth,
and $\mathcal{K}=n_s\epsilon_{kin}$, we find
\beq
\mathcal{K}=\frac{m_e}{m^*} \frac{H^2}{8\pi}
\eeq
which disagrees with Eq. (57) and is incorrect.  This shows once again that expressions (21a) for the superfluid velocity
and Eq. (19) for the London penetration depth in terms of the effective mass rather than the bare electron mass are
untenable.

\section{Experimental determination of the superfluid velocity}

            \begin{figure}
 \resizebox{4.5cm}{!}{\includegraphics[width=6cm]{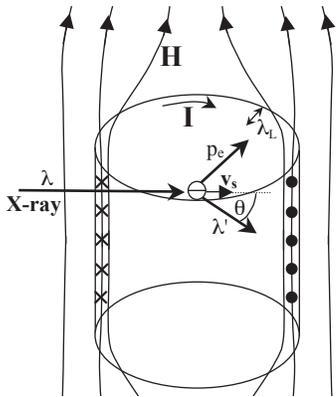}}
 \caption { Compton scattering experiment on a superconducting cylinder in a magnetic field.
 The wavelength of the scattered X-ray for given scattering angle $\theta$ will be shifted proportionally to the superfluid velocity.  }
 \label{figure1}
 \end{figure} 
 
 Compton scattering experiments \cite{compton1,compton2} offer a straightforward way, at least in principle, to measure the superfluid speed  
 \beq
v_s=-\frac{e\lambda_L}{m_e c}H
\eeq
in a superconductor, as shown in Fig. 2. This is simply because in Compton scattering 
within the impulse approximation an individual photon
is scattered by an individual electron, so the superfluid density does not play a role in determining
the final state of the photon.
For an X-ray incident parallel   to the direction of the superfluid velocity, the difference between scattered and incident
 wavelengths of the photon for photon scattering angle $\theta$ is simply
 \beq
 \lambda'-\lambda=\lambda_c (1 \pm  \frac{v_s}{c}(1+\frac{\lambda}{\lambda_c}))(1-cos\theta)
 \eeq
 where the $+$/$-$ corresponds to the superelectron moving in the same / opposite direction to the incident photon. 
 $\lambda_c=h/m_ec$ is the Compton wavelength. A typical value
 for the superfluid speed Eq. (72) for $\lambda_L=400A$ and $H=500G$ is $v_s=35,225cm/s$. If we take for the applied magnetic field
 \beq
 H=\frac{\hbar c}{4e\lambda_L^2}
 \eeq 
 which is approximately the lower critical field $H_{c1}$ \cite{tinkham1}, the superfluid speed is
\beq
 v_s=\frac{\hbar}{4 m_e \lambda_L}
 \eeq
 and Eq. (73) takes the simple form
 \bmath
 \beq
  \lambda'-\lambda=\lambda_c (1 \pm  \frac{\lambda+\lambda_c }{8\pi \lambda_L})(1-cos\theta) .
  \eeq
  If instead of Eq. (72) we assume the superfluid speed is given by the BCS formula Eq. (21a)
  we obtain instead
   \beq
  \lambda'-\lambda=\lambda_c (1 \pm  \frac{m_e}{m^*}\frac{\lambda+\lambda_c }{8\pi \lambda_L})(1-cos\theta) .
  \eeq
  \emath

  In the absence of supercurrent, the Compton scattering profile will be Doppler broadened by the velocity of the Bloch electrons.
  In the presence of the supercurrent the velocity  of any given electron in the supercurrent will be
  \beq
  \vec{v}_k'=\vec{v}_k+ \vec{v}_s
  \eeq
  with $\vec{v}_k$ its velocity in the absence of supercurrent.
  Thus, the Compton profile for an incident monochromatic beam will be   shifted  by the amount 
  given by Eq. (72).
  Even though the shift is much smaller than the Doppler broadening, because it is the same for all scattering angles
  $\theta$ it will hopefully be detectable with currently available resolution by accumulating measurements for 
  many angles. Thus a quantitative measurement of this Compton shift
  together with an independent measurement of $\lambda_L$ should be able to prove (or disprove) experimentally that it is
  the bare electron mass that enters Eq. (72) for the superfluid speed rather than the effective mass.

  In addition, Compton scattering can check the prediction of the theory of hole superconductivity \cite{holesc}  that a spin current
  exists in the ground state of superconductors that flows within a London penetration
  depth of the surface in the absence of applied fields \cite{electrospin}. In the geometry of Fig. 2, the supercarrier of spin $\vec{\sigma}$ 
  is predicted to have 
 azimuthal  velocity
   \beq
 \vec{v}_{\sigma}  =-\frac{\hbar}{4 m_e \lambda_L}\vec{\sigma}\times \hat{n} -\frac{e\lambda_L}{m_e c}\hat{n}\times\vec{H}
\eeq
where $\hat{n}$ is the outward pointing normal to the lateral surface of the cylinder. The carriers moving clockwise,
with spin $\vec{\sigma}//\vec{H}$,
are brought to a stop when the magnetic field reaches the value Eq. (74). For $H=0$, carriers of opposite spin flow in
opposite direction with speed given by Eq. (75).  Through spin-dependent Compton scattering with circularly polarized photons \cite{compton3}
it will hopefully be possible to verify or disprove this prediction.

 \section{Discussion}
 
 In this paper we have pointed out that the speed of the supercurrent 
 carriers in superconductors $v_s$, or equivalently the effective mass of carriers of the supercurrent $m^*$, or equivalently the 
 density of superfluid carriers $n_s$, can be measured experimentally. This is contrary to the generally accepted view that only
 the combination $n_s/m^*$ can be directly measured, and the related  generally accepted
 view that the superfluid
 velocity $\vec{v}_s$ is not a physical observable, only the current density $\vec{J}$ given by Eq. (21b) is. Measurement 
 of the superfluid velocity should  be able to confirm the theoretical claim discussed in
this paper. However, even if the experimental accuracy required will not be attainable in the
foreseeable future, it is useful to think of it as a `gedankenexperiment' that can $in$ $principle$
determine what the value of the superfluid speed $v_s$ is. 
The fact that $v_s$ is a physical observable allows us to make the theoretical claim that  {\it BCS theory applied to Bloch electrons is internally inconsistent}. In the following we summarize our arguments.
In essence, we claim that BCS theory cannot describe supercurrent flow of carriers that
have an effective mass $m^*$ in the normal state that is different from the bare electron mass $m_e$,
and at the same time be consistent with a wide range of experimental properties of superconductors.

Before we start we should address the predictable objection of critical readers that we have ignored a large number
of factors that will modify the effective  mass besides the electron-ion interaction,  as well as other effects not considered here: Fermi liquid effects, non-Fermi-liquid effects,
electron-phonon interactions, renormalizations, multiband effects, long-range Coulomb interactions, 
spin-orbit interactions, magnetic interactions, Kondo physics, disorder, topological effects, 
relativistic effects, etc. 
In a real material all these effects may play a role. Nevertheless, we argue that it is 
a valid and useful theoretical approach to establish the inconsistency of BCS theory and band theory of solids
assuming all these other effects can be ignored. 
While it is not $impossible$ that including some or all of these other effects could restore the
consistency of BCS theory we don't see a shred of a hint for why this would be the case. In any event it is
a matter for the future to decide.

In this paper we assume electrons interact with ions as described by the standard Bloch theory
of electronic energy bands in solids, disorder can be ignored, and electrons
behave as independent particles except for a weak attractive interaction that leads
to the BCS superconducting state below a critical temperature. 
We furthermore assume only one band is partially filled and hence conducts electricity,
all other bands are either full or empty. For simplicity we have assumed in Appendix A that
the partially filled band is the lowest band, but this restriction can be removed without 
altering our arguments and conclusions. We assume the system is at zero temperature.
 
Consider the following five points that we argue are well established:
 
 (1) Using semiclassical transport theory (Sect. II) or equivalently the standard Kubo linear response formalism
 (Appendix A) it follows that within BCS theory or London theory the speed of electrons in the supercurrent is given by Eq. (21a), involving the band mass
 $m^*$ rather than the bare electron mass $m_e$, for the cases when a band is nearly empty or
 the band is nearly full. In the first case, the density $n_s$ of carriers carrying  the supercurrent
 at zero temperature is the total electron density, in the second case it is the total hole density.
 
 (2) The mechanical momentum of the carriers carrying the supercurrent is the product of the $bare$
 electron mass $m_e$ and the superfluid velocity:
 \beq
 \vec{p}_{mech}=m_e \vec{v}_s  .
 \eeq
 This follows theoretically from the semiclassical treatment
 or equivalently from linear response theory, and is quantitatively verified by 
 gyromagnetic effect experiments.
 
(3) The kinetic energy density of the supercurrent is given by
 \beq
\mathcal{K}=\frac{H^2}{8\pi}  .
\eeq
 This follows from general properties of the superconducting state.
 
 (4) Superconductors exhibit many properties that establish that the superfluid is described
 by a single-particle-like macroscopic wavefunction $\Psi(\vec{r})$ that has
 a well-defined macroscopic phase $\varphi(\vec{r})$. The gradient of the phase is
 related to the superfluid velocity and the magnetic vector potential.
 
 (5) For a rotating superconductor, the superfluid speed is given by Eq. (38) (the first equality) involving the bare electron mass,
 both in the interior region where the superfluid moves together with the body and near the surface where the superfluid
 lags the motion of the body \cite{londonbooklm}.

 We believe that the points (1)-(5) above are generally accepted, well established,
 experimentally proven, and true. Our claim is that BCS theory can be made
 to be consistent with some of those points, but not with all.
 
To start, we need to decide what is the equation relating the gradient of the phase, the superfluid 
velocity and the magnetic vector potential. We argue that Eq. (30) with the effective mass $m^*$
or Eq. (37) with the bare mass $m_e$ are the only reasonable choices. Then we argue that
with either choice BCS theory is internally inconsistent. We discuss both choices in turn.

If Eq. (37) for the phase is valid, it leads to Eq. (38), the speed of the supercurrent, depending on $m_e$ rather than
$m^*$. This is consistent with experiments where the entire body moves
and the speed $v_s$ describes both the speed of the superfluid and the 
speed of the body \cite{hild,zimm,parker}, however it is inconsistent when the body is at rest and
a supercurrent flows. It  can only be made compatible with BCS assuming the density of
supercurrent carriers is neither given by the density of electrons for a nearly empty band, nor
by the density of holes for a nearly full band, as shown in Appendix B. 
This assumption would lead to the conclusion that a superconductor and a 
perfect conductor respond differently to an applied magnetic field.
This is inconsistent with our general understanding of superconductivity.
It would also be inconsistent with BCS theory at finite temperatures,
and would require that not all the Cooper pairs contribute to the supercurrent,
in contradiction with BCS theory.

 If instead Eq. (30) for the phase is valid, it is consistent with the Meissner effect when
 the phase is constant, with the superfluid velocity given by the BCS expression
 Eq. (21a) involving the effective mass.
 However, Eq. (30) requires that the mechanical momentum of electrons in the
 supercurrent is
  \beq
 \vec{p}_{mech}=m^* \vec{v}_s
 \eeq
 and this is inconsistent with Eq. (79) for the mechanical momentum of supercurrent carriers, 
 which follows from
 BCS theory $and$ agrees with experiment.
 
 Eq. (30) for the phase is also incompatible with experiments where the 
 entire body moves  \cite{hild,zimm,parker}. 
 One could try to argue that Eq. (37) should be used when the 
 superfluid moves together with the body, and Eq. (30) should be used when
 the superfluid moves and the body is at rest \cite{scalsuggestion}.  However,
 it is not clear then what should be used when both the body is moving
 and the superfluid is moving relative to the body as in the Parker-Simmonds experiment \cite{parker}.
Even for the rotating superconductor where the superfluid rotates together with the body in the interior, in the
region within a London penetration depth of the surface there is  relative motion 
of the supercurrent and the body to generate the interior magnetic field $\vec{B}_i=-(2m_e c/e) \vec{\omega}$ \cite{londonbooklm}. Consider what happens if we apply an external
magnetic field $\vec{H}=\vec{B}_i$ to the rotating superconductor. Electrons will respond to the Faraday electric field by changing their
speed according to Eq. (21a) with $m^*$. However if the governing phase equation for the rotating superconductor is Eq. (37), the drift velocity
before $\vec{H}$ was applied was Eq. (38) with $m_e$. This would imply that the drift current does not stop, which is unphysical.
Physically we expect 
that the surface drift current will stop resulting in a uniform magnetic field $\vec{B}_i$ both inside and outside the superconductor.

 In addition, we have shown in Sect. VI that BCS theory
 in the standard many-body treatment of Appendix A is incompatible
 with the phase rigidity that is implied by either the phase equations
 Eq. (30) or Eq. (37). As a consequence,
 BCS theory does not describe the irrotational superfluid flow required by the 
 phase equations that is characteristic of
 superfluids, and it cannot predict the flux quantization for a 
 multiply connected sample that is predicted by the phase equations and observed experimentally.
 The BCS wavefunction is not rigid, contrary to what the phase equations imply,
 rather it is modified by a long wavelength magnetic field because the 
 perturbation induces transitions between electrons in the band 
 responsible for superconductivity and other bands, as
 shown in Appendix A.
 
 Finally and independently, we have argued that BCS theory is incompatible with the known
 expression for the kinetic energy density of the supercurrent Eq. (80).
 This expression requires that the kinetic energy of a superfluid carrier is
 \beq
\epsilon_{kin}=\frac{1}{2}m^* v_s^2  .
\eeq
 However this is not the kinetic energy of electrons within
 Bloch theory of solids under any circumstances. Using $m_e$ as prefactor in Eq. (82)
 is the correct expression for the kinetic energy of Bloch electrons 
  under the assumption that the potential energy of 
carriers is independent of $v_s$, however this would lead to the kinetic energy density
being given by Eq. (71) which is different from Eq. (80) and hence incorrect.
 
 In summary, we have given several different independent arguments that
 establish that BCS theory to describe superconductivity in nature and Bloch theory of solids are mutually
 incompatible. The conclusion is that {\it BCS theory in its current form is only
 consistent if we assume it applies to a free electron system}, i.e. if $m^*=m_e$.

 When F. London first introduced Eq. (30) \cite{london1948}, written in the form
 \beq
 \vec{p}_s=(\frac{m^*}{n_se})\vec{J}+\frac{e}{c}\vec{A}
 \eeq
he called the left-hand-side
 the `mean momentum field of
the superelectrons' and pointed out that $\vec{\nabla}\times\vec{p}_s=0$ describes the Meissner effect \cite{london1948}. He assumed that $\vec{p}_s$ was the same canonical momentum that appears in the
 Schr\"odinger equation, $\vec{p}=(\hbar/i)\vec{\nabla}$, and pointed out that in the normal state
  it adopts the `local value' $(e/c)\vec{A}$ in the presence of a magnetic vector
 potential to minimize the kinetic energy $(\vec{p}-(e/c)\vec{A})^2/(2m_e)$,
 while in a superconductor it is prevented from doing so because of `rigidity' of the wavefunction.
 However, he failed to notice (or to point out) that the first term on the right-hand-side of Eq. (83) is $not$ the
 mechanical momentum of the superelectrons if it involves $m^*$ rather than $m_e$, which converts
 Eq. (83) into a completely ad-hoc Ansatz with no relation to the Schr\"odinger equation that
 ultimately governs the behavior of the microscopic components of a superconductor. This may perhaps
 be termed the `original sin' from which the contradictions discussed in this paper originated.

The findings discussed in this paper
 imply  that the conventional BCS-London theory of superconductivity applied to electrons in energy bands of solids, as done in 
 Refs. \cite{einzel,einzel2,londonours,scal,misawa,carbotte,chak,franz,gross,tinkham,londonbook,london1948,laue1948,schriefferbook} and innumerable others, 
is   {\it internally inconsistent} as well as  inconsistent with well-established experimental properties of
  superconductors.
  This conclusion has wide-ranging implications:   there are many simple metals believed to be BCS superconductors described
  by BCS theory \cite{spissue}, and there are many normal state properties of metals that are rather well explained by Bloch's band
  theory of solids \cite{am}. 
Faced with these facts, what is the way out of this conundrum?  
    
  The problem, we propose, lies  in the key BCS assumption that the states that define the BCS wavefunction Eq. (7) are the
  same Bloch states as in the normal state. This is properly recognized to be an assumption in Ref. \cite{einzel}. The BCS amplitudes $\bar{u}_k$ and $\bar{v}_k$ in Eq. (7) \cite{tinkham} differ from
  their values in the normal state only for values of $\epsilon_k$ within a region of width $\Delta$ of the Fermi energy, where $\Delta$ is the BCS energy gap. For `conventional' superconductors this is certainly a tiny fraction of all the conduction
  electrons (or holes) in the band. Within BCS all the dramatic changes in the properties of a metal that
  undergoes a transition to the superconducting state result from a 
  redistribution of the occupation of these Bloch states in the superconducting state, and all the
  other conduction electrons in the system, which is the vast majority, are unaffected. This is a rather remarkable
  statement, that condensed matter physicists have adhered to for the last 60 years. What if it is not true?

  The considerations in this paper suggest that the only consistent way to interpret experiments in superconductors is
  to assume that electrons in the superconducting state of metals respond to applied external fields as
  {\it perfectly free electrons}. In other words, that carriers condensing into the superfluid state and
  contributing to the supercurrent 
become  completely `undressed' from   electron-ion,
  electron-electron and electron-phonon interactions, that in the normal state dress the electron and make it respond with an effective
  mass $m^*$ rather than its bare mass $m_e$. This assumption consistently explains the experimental observations discussed
  in this paper, and it says that Eq. (37) for the phase is valid and Eq. (30) is invalid. 
  It also explains why the work done by the external field changes only the kinetic energy and not the
  potential energy of superfluid carriers  as assumed in Sect. VII.
  
  We have assumed an isotropic band structure in this paper for simplicity. Of course many real solids are anisotropic.
  In the conventional treatment it is assumed that the effective mass is a tensor \cite{laue1948}. Here we propose  instead that
  the mass is a scalar, the free electron mass, and anisotropies are described by a carrier density tensor, i.e.
  that the relation between current density and magnetic vector potential is
  \bmath
  \beq
  \vec{J}=-\frac{e^2}{m_ec } \overleftrightarrow{n_s} \cdot \vec{A}
  \eeq
  and the London kernel is
  \beq
   \overleftrightarrow{K_L}=\frac{4\pi e^2}{m_e c^2} \overleftrightarrow{n_s}
  \eeq
  \emath
  so that along a principal axis the current is given by
  \beq
  J_i=-\frac{e}{m_e c}n_s^{ii}A_i=n_s^{ii} e (\vec{v}_s)_i
  \eeq
with  the superfluid velocity  given by
  \beq
  \vec{v}_s=-\frac{e}{m_e c}\vec{A} ,
  \eeq
and the superfluid mechanical momentum for one carrier in a simply connected superconductor is  given by
\beq
\vec{p}_{mech}=m_e\vec{v}_s=-\frac{e}{c}\vec{A}   .
\eeq
It follows from the discussion in this paper that whether to ascribe observed anisotropies in the London
penetration depth to an effective mass tensor or to a carrier density tensor is $not$ semantics, as generally assumed.
The latter is the $only$ possible choice, and it is experimentally verifiable through Compton scattering experiments.

Assuming readers agree that the points made in this paper are correct, and even before experimental confirmation
by Compton scattering experiments, we suggest that the focus of theoretical research in superconductivity should switch to
understanding how normal carriers in solids, governed by complicated band structures and `dressed' by electron-ion, electron-electron
and electron-phonon  interactions, become completely `undressed' from these interactions so that they respond as free electrons in the superconducting state. This was in fact the generally held 
view in the early days of superconductivity \cite{becker}.  

Within the alternative theory of hole superconductivity \cite{holesc}, carriers in a nearly filled band  are highly dressed in the normal
state  \cite{holeundr,ehasym},
and when going superconducting   they expand their wavelength \cite{sm}  so that they no longer `see' the lattice periodic potential, 
hence `undress' from the electron-ion interaction \cite{holeelec2}  and respond as free electrons. We suggest that an answer to the questions posed in this paper may
be found along those lines.
 
\appendix

\section{Calculation of the London kernel}

In this appendix we calculate the supercurrent  and the mechanical momentum density using the standard linear response formalism and show 
that they agree with the results obtained in Sect. II and III. 

In first quantized form, the electric current density is given by $\vec{J}=(e/V)\sum_i\vec{v}_i$, with V the volume
and the velocity of the i-th particle given by
\beq
\vec{v}_i=\frac{1}{m_e}(\vec{p}_i-\frac{e}{c}\vec{A}(\vec{r}_i))
\eeq
with $\vec{p_i}$ the canonical momentum operator for the i-th particle
 $(\vec{p}_i=(\hbar/i) \vec{\nabla}_i) $, so the current density is
\beq
\vec{J}=\frac{e}{m_e V}\sum_i\vec{p}_i-\frac{e^2}{m_e c V} \sum_i\vec{A}(\vec{r}_i)\equiv \vec{J}_1+\vec{J}_2
\eeq
with $\vec{J}_1$ and $\vec{J}_2$ the so-called  paramagnetic and diamagnetic currents.
We assume the vector potential $\vec{A}$ is in the Coulomb gauge, $\vec{\nabla}\cdot \vec{A}=0$. 
Next we rewrite $\vec{J}_1$ and $\vec{J}_2$ in second quantized form using as single
particle basis the Bloch eigenfunctions of the single electron problem in the lattice ionic potential:
\beq
\Psi_{nk}(\vec{r})\equiv<\vec{r}|nk>
\eeq
with $n$ labeling the n-th band and $k$ crystal momentum (we omit vector notation on k  for simplicity) to obtain
\bmath
\beq
\vec{J}_1=\frac{e}{m_eV} \sum_{n,n',k,\sigma}c_{nk\sigma}^\dagger c_{n'k\sigma}
<nk|\vec{p} |n'k>
\eeq
\beq
\vec{J}_2=-\frac{e^2}{m_e c}\frac{1}{V}( \sum_{nk\sigma}c_{nk\sigma}^\dagger c_{nk\sigma}) \vec{A}
\eeq
\emath
assuming a uniform vector potential $\vec{A}$.  The operator $c_{nk\sigma}^\dagger$ creates an electron with 
wavefunction $\Psi_{nk}(\vec{r})$ and spin $\sigma$ in the n-th band,  with band energy $\epsilon_{nk}$.

We  assume for simplicity an isotropic system 
and zero temperature.
To lowest order in $\vec{A}$ the currents are given
by
\beq
\vec{J}=\vec{J}_1+\vec{J}_2=-\frac{c}{4\pi}(K_1+K_2)\vec{A}
\eeq
with $K_1$ and $K_2$  the  so-called   `paramagnetic' and
`diamagnetic' London kernels. $K_2$ is simply obtained by taking the expectation value of $\vec{J}_2$
in the BCS ground state
\beq
|\Psi_G>=\prod_k (\bar{u}_k+\bar{v}_k c_{0k\uparrow}^\dagger c_{-0k\downarrow}^\dagger) |0>
\eeq
where the partially filled band $n=0$ is the band giving rise to superconductivity. We assume for simplicity this is the
lowest energy band, so that all other bands $n>0$ are empty at zero temperature in the absence of
applied fields. To lowest order in $\vec{A}$, the diamagnetic current is simply the expectation value of
$\vec{J}_2$ in the BCS ground state:
\beq
<\Psi_G|\vec{J}_2|\Psi_G>=-\frac{e^2}{m_e c}\frac{1}{V}\sum_k 2|\bar{v}_k|^2 \vec{A}
\eeq
so that the diamagnetic kernel is
 \beq
K_2=\frac{4\pi e^2}{m_e c^2}( \frac{1}{V}\sum_k n_k) .
\eeq
with $n_k=2|\bar{v}_k|^2 $ as given by Eq. (16).

To compute the paramagnetic kernel we need the ground state wavefunction to first order in $\vec{A}$. 
The perturbing Hamiltonian in first quantized form is given by expanding
$(\vec{p}-(e/c)\vec{A})^2/2m_e$ as
\beq
H_1=-\frac{e}{2m_e c}\sum_i(\vec{p}_i\cdot\vec{A}(\vec{r}_i)+\vec{A}(\vec{r}_i)\cdot\vec{p}_i)
\eeq
Its second quantized form for uniform $\vec{A}$ is  
\beq
H_1=-\frac{e}{m_e c}\   [\sum_{n n' k \sigma} 
c_{nk\sigma}^\dagger c_{n'k\sigma}
[<nk|\vec{p} |n'k>]] \vec{A}
\eeq
The wavefunction to first order in $\vec{A}$ is given by
\beq
|\Psi>=|\Psi_G>-\sum_m\frac{<\Psi_m|H_1|\Psi_G>}{E_m-E_G}|\Psi_m>
\eeq
where $E_G$ is the energy of the BCS ground state, and $E_m$ the energy of the excited
state $|\Psi_m>$.
Taking the expectation value of the paramagnetic current $\vec{J}_1$ with this wavefunction yields
\beq
<\Psi|\vec{J}_1|\Psi>=-2Re[\sum_m\frac{<\Psi_G|\vec{J}_1|\Psi_m><\Psi_m|H_1|\Psi_G>}{E_m-E_G}]
\eeq

When $H_1$ operates on the BCS ground state it destroys an electron in band $n'=0$ and creates one
in band $n$, either $n=0$ or $n>0$. Thus, there are two types of contributions to Eq. (A12) resulting from
the states with $n=0$ and $n>0$ respectively. For the contribution from the   $n=0$  states
the calculation is exactly as described in Tinkham \cite{tinkham}, and yields zero at zero temperature
for a uniform vector potential. Thus the only contributions to Eq. (A12) come from excited states where
there is one electron in a band $n>0$. Eq. (A12) then yields
\beq
<\Psi|\vec{J}_1|\Psi>=\frac{4e^2}{m_e^2cV}\sum_{n>0,k}
|\bar{v}_k|^2 \frac{<0k|p|nk><nk|p|0k>}{\epsilon_{nk}+E_k-\mu}\vec{A}
\eeq
with
\beq
E_k=\sqrt{(\epsilon_{0k}-\mu)^2+\Delta^2}
\eeq
the BCS quasiparticle excitation energy, $\mu$ the chemical potential and $\Delta$ the BCS gap.
We are assuming an isotropic system and $p$ is any one component of the momentum operator.

Now the oscillator strength sum rule for Bloch electrons for isotropic energy bands yields \cite{wooten}
\beq
 \frac{2 }{m_e^2}\sum_{n\neq 0}  \frac{<0k|p|nk><nk|p|0k>}{\epsilon_{0k}-\epsilon_{nk}}
 =\frac{1}{m^*_k}-\frac{1}{m_e}
 \eeq
 with $m^*_k$ defined by Eq. (12). The sum rule results from   expanding $\epsilon_{0,k+q}$ to second order in q and using second order perturbation theory. We can use Eq. (A15) in Eq. (A13) if we approximate
 \beqn
 \epsilon_{nk}+E_k-\mu&=& \epsilon_{nk}+\sqrt{(\epsilon_{0k}-\mu)^2+\Delta^2}-\mu  \nonumber \\
 &\sim& \epsilon_{nk}-\epsilon_{0k}
 \eeqn
 which should be an excellent approximation when $|\bar{v}_k|^2\neq 0$. Eq. (A13) then yields
 \beq
 <\Psi|\vec{J}_1|\Psi>=-\frac{e^2}{cV}\sum_k 2|\bar{v}_k|^2(\frac{1}{m^*_k}-\frac{1}{m_e})\vec{A}
 \eeq
 and we obtain for the paramagnetic London kernel 
  \beq
 K_1=\frac{4\pi e^2}{c^2}\frac{1}{V}\sum_k n_k(\frac{1}{m^*_k}-\frac{1}{m_e})
 \eeq
 and for the total kernel
  \beq
 K=K_1+K_2=\frac{4\pi e^2}{c^2}\frac{1}{V}\sum_k n_k\frac{1}{m^*_k}
 \eeq
 in agreement with Eq. (18). The total current  to first order in $\vec{A}$ is
 \beq
<\Psi|\vec{J}_1+\vec{J}_2|\Psi>=-\frac{e^2}{cV}\sum_k n_k(\frac{1}{m^*_k})\vec{A}
\eeq
which agrees with Eq. (17) since $A=\lambda_LH$ in the cylindrical geometry under consideration.

The same formalism applies to the mechanical momentum density $\mathcal{P}_{mech}$. From Eq. (A1),
the mechanical momentum of a carrier is
 \beq
\vec{p}_i^{mech}=m_e \vec{v}_i= \vec{p}_i-\frac{e}{c}\vec{A}(\vec{r}_i)
\eeq
and the mechanical momentum density is 
\beq
\vec{\mathcal{P}}_{mech}=
\frac{1}{V}\sum_i\vec{p}_i-\frac{e}{ c V} \sum_i\vec{A}(\vec{r}_i)\equiv \vec{\mathcal{P}}_1+\vec{\mathcal{P}}_2
=\frac{m_e}{e}(\vec{J}_1+\vec{J}_2)
\eeq
The `paramagnetic' and `diamagnetic' momentum densities to first order in $\vec{A}$ are
\bmath
  \beq
 <\Psi|\vec{\mathcal{P}}_1|\Psi>=-\frac{m_e e}{cV}\sum_k n_k(\frac{1}{m^*_k}-\frac{1}{m_e})\vec{A}
 \eeq
\beq
<\Psi_G|\vec{\mathcal{P}}_2|\Psi_G>=-\frac{e}{c}\frac{1}{V}\sum_k n_k \vec{A}
\eeq
\emath
and the total mechanical momentum density is
  \beq
 <\Psi|\vec{\mathcal{P}}_{mech}|\Psi>=-\frac{m_e e}{cV}\sum_k n_k\frac{1}{m^*_k}\vec{A}=-\frac{m_e n_s e}{m^*c}\vec{A}
 \eeq
 in agreement with Eq. (25) since $A=\lambda_L H$. 
 
 Note that $\mathcal{P}_1$ is the canonical momentum density for the many-electron system.
 The fact that its $T=0$ expectation value Eq. (A23a) is non-zero to first order in $\vec{A}$ indicates
 that the BCS wavefunction is not `rigid' with respect to magnetic perturbations, 
 contrary to what is generally believed.
 
 \section{Concise formulation of the BCS inconsistency in terms of the superfluid density}
 In this appendix we discuss one aspect of the inherent inconsistency of BCS theory in terms that some readers
 may find more appealing.
 
 Assuming the validity of Eq. (37) \cite{scal2,zimm,jak,parker,hild} so that the mechanical momentum
 is correctly given by $m_e v_s$, we have for a simply connected superconductor that   $\vec{\nabla}\varphi=0$ and hence
 \beq
\vec{v}_s=-\frac{e}{m_ec}\vec{A}
\eeq
for the velocity of Cooper pairs.
Calling  $n_s/2$ the number of Cooper pairs per unit volume, each with charge 2e, the supercurrent density is then
\beq
\vec{J}=n_se\vec{v}_s=-\frac{e^2}{c}\frac{n_s}{m_e} \vec{A}
\eeq

The BCS wave function   is given by Eq. (7), and $m^*_k$ is defined in Eq. (12).
The current density   to first order in  $\vec{A}$ at zero temperature is, 
as shown in Appendix A, Eq. (A20) 
 \beq
\vec{J}=-\frac{e^2}{c}\frac{1}{V}\sum_k 2|\bar{v}_k|^2(\frac{1}{m^*_k})\vec{A}
\eeq
 which also applies to tight binding models such as the  Hubbard model.
Comparing Eqs. (B2) and (B3),
\beq
\frac{n_s}{m_e}=\frac{1}{V}\sum_k 2|\bar{v}_k|^2(\frac{1}{m^*_k})
\eeq
The density of electrons in the system described by the BCS wavefunction Eq. (7) is
\beq
n_e=\frac{1}{V}<\Psi|(\sum_{k\sigma} c_{k\sigma}^\dagger c_{k\sigma})|\Psi>=\frac{1}{V}\sum_k 2|\bar{v}_k|^2
\eeq
When  solving for the BCS wavefunction one  picks the chemical potential $\mu$ so that $n_e$ given by Eq. (B5) yields the density of electrons in the normal state. $\mu$ will be very close to the Fermi energy in the normal state. 

For a band close to empty we will have $m^*_k\sim m^*$ for the occupied states independent of $k$, hence from Eqs. (B4) and (B5)
\beq
n_s=(\frac{m_e}{m^*})n_e
\eeq

Similarly for a band close to full we will have $m^*_k\sim (-m^*)$ independent of $k$ for the empty states, the density of empty states in the band is
\beq
n_h=\frac{1}{V}\sum_k 2(1-|\bar{v}_k|^2)
\eeq
and
\beq
n_s=(\frac{m_e}{m^*})n_h
\eeq
These equations are valid in the absence of disorder at zero temperature. In the presence of disorder they
will remain essentially unchanged in the clean limit \cite{mars}.

Eqs. (B6) and (B8) say that the number of carriers in the supercurrent, $n_s$, at zero temperature, in the clean limit, is not equal to the number of electrons in the Bloch or tight binding band when the band is almost empty, nor equal to the number of holes when the band is almost full. An explanation of this inconsistency has not been
proposed in the scientific literature to our knowledge. It implies that 
a perfect conductor and a superconductor would behave differently under application of a
magnetic field, which is contrary to the general understanding.
Note that it resembles what has been termed the
`condensate saga' in the study of superfluid $^4He$ \cite{saga}, the fact that the measured condensate
fraction is significantly lower than the superfluid $^4He$ density \cite{platz}.

\acknowledgements
The author is grateful to F. Marsiglio for helpful discussions. He is grateful to D. Scalapino for 
calling Ref. \cite{parker} to his attention and for many stimulating discussions, in particular
on the relevance of Ref. \cite{jak}, and for
suggesting that in his view Eq. (30)  applies when the solid is at rest and Eq. (37) applies when the
solid moves with the superfluid.   He is also grateful to 
D. Einzel for comments and sharing his unpublished notes, and to 
P. Hirschfeld, Congjun Wu, L. Sham, S. Sinha, A. Leggett and N. Goldenfeld for comments and  their interest in this work.


\begin{references}
\bibitem{footnote}
The reader may argue that the problem can be fixed if instead of Eq. (31) the superfluid wavefunction is given by
\beq
\Psi_S(\vec{r})=(n_s^*/2)^{1/2}e^{i\varphi(\vec{r})} \nonumber
\eeq
with $n_s^*=n_s m_e/m^*$. If so the correct mechanical momentum density results from Eq. (36).
However, the current density carried by this wavefunction   would be
\beq
\vec{J}=2e|\Psi_S|^2 \vec{v}_s=n_s^* ev_s \nonumber
\eeq
which is different from Eq. (3) and hence incorrect.
\end{references}
\end{document}